\documentstyle[12pt]{article}
\begin{document}
\begin{center}
{\bf{\large{Non static local string in Brans-Dicke theory}}}\\
\vskip 20pt

A.A.Sen\footnote{E-mail:anjan@juphys.ernet.in}\\

Relativity and Cosmology Research Centre\\
Department of Physics, Jadavpur University\\
Calcutta 700032, India.\\
\end{center}

\centerline{\underline{Abstract}}

A recent investigation shows that a local gauge string with a 
phenomenological energy momentum tensor, as prescribed by Vilenkin, is 
inconsistent in Brans-Dicke theory. In this work it has been shown that
such a string is indeed consistent if one introduces time dependences 
in the metric.
A set of solutions of full nonlinear Einstein's equations for the interior 
region of such a string are presented.
\newpage

The  importance of nonminimally coupled scalar tensor theories such as Brans-Dicke
theory, in the context of inflationary scenario has generated a considerable 
interest in the gravitational field of certain topological defects, 
particularly cosmic strings in such theories \cite{R1}, \cite{R2}, \cite{R4},
\cite{R5}, \cite{R6}. Very recently it has been shown by the present authors 
that a local static gauge string, given by the energy momentum tensor 
components $T^{t}_{t}=T^{z}_{z}\neq 0$ and all other $T^{\mu}_{\nu}=0$ 
\cite{R7} is inconsistent in B-D theory of gravity \cite{R8}, although it is
quite consistent in a more general scalar tensor theory \cite{R9}, where
the parameter  $\omega$ is a function of the B-D scalar field.

A very recently communicated investigation by Dando and Gregory \cite{R10}
has shown that one can have a nonsingular spacetime for a global string
in dilaton gravity if one includes time dependences in the line element. In
this work, we have shown in a very similar manner that one can obtain a 
consistent set of nonstatic solution for local gauge string with
$T^{\mu}_{\nu}$ mentioned above in B-D theory.

The gravitational field equations in B-D theory are given by
 \begin{equation}
 G_{\mu\nu}= {T_{\mu\nu}\over{\phi}}+{\omega\over{\phi^{2}}}(\phi_{,\mu}\phi_{,\nu}-
 {1\over{2}}g_{\mu\nu}\phi^{,\alpha}\phi_{,\alpha})+{1\over{\phi}}(\phi_{,\mu;\nu}-
 g_{\mu\nu}\Box\phi),
 \end{equation}
 where $\omega$ is a dimensionless constant parameter of the theory  
 and $\phi=\phi(r)$ is the B-D scalar field.

 The wave equation for the scalar field $\phi$ is 
 \begin{equation}
 \Box\phi={T\over{(2\omega+3)}}
 \end{equation}
 In these equations, $T_{\mu\nu}$ represents the energy momentum 
 tensor components for all the fields except the scalar field $\phi$ 
 and $T$ is the trace of $T_{\mu\nu}$.

 We take the line element describing a cylindrically symmetric spacetime,
 with two killing vectors $\partial_{z}$ and $\partial_{\phi}$ as in \cite{R11},
 $$
 ds^{2}=e^{2A(r)}[dt^{2}-e^{2B(t)}dz^{2}]-D^{2}(t)[dr^{2}+C^{2}(r)d\theta^{2}],
 \eqno{(3)}
 $$
 where $r, \theta, z$ are cylindrical coordinates defined in the range
 $0\leq r \leq \infty; 0\leq\theta\leq\infty; -\infty\leq z\leq\infty$.

 The local gauge string is  characterised by an energy density and a stress 
 along the symmetry axis, given by \cite{R7}
 $$
 T^{t}_{t}=T^{z}_{z}= \sigma(r)
 \eqno{(4a)}
 $$
 $$
 T^{r}_{r}=T^{\theta}_{\theta}=0.
 \eqno{(4b)}
 $$

 The conservation of matter is represented by the equation
 $$
 T^{\mu\nu}_{;\nu}=0.
 \eqno{(5)}
 $$
 It should be noted that (5) and (2) are not independent equations, as in 
 view of (1) and the Bianchi identity, one yields the other.

 Eqn (5) together with (4a), (4b) and (3) yields
 $$
 {\dot{D}}=0,
 \eqno{(6a)}
 $$
 $$
 A^{\prime}=0
 \eqno{(6b)}
 $$
 for $\sigma \neq 0$. Here an over head dot and prime represent 
 differentiations w.r.t t and r respectively.

 So $D$ and $A$ are constants and in what follows we shall take $D=1$ and 
 $e^{2A}=1$ which leads only to a rescaling of coordinates and no loss 
 generality.

 In view of (6a) and (6b), the field equations (1) can be written as,
 $$
 {C^{\prime\prime}\over{C}}=-{\sigma\over{\phi}}-{\omega\over{2}}
 {\phi^{\prime 2}\over{\phi^{2}}}-{\phi^{\prime\prime}\over{\phi}}-
 {\phi^{\prime}\over{\phi}}{C^{\prime}\over{C}}
 \eqno{(7a)}
 $$
 $$
 \ddot{B}+{\dot{B}}^{2}=  -{\omega\over{2}}{\phi^{\prime 2}\over{\phi^{2}}}   
 + {\phi^{\prime}\over{\phi}}{C^{\prime}\over{C}}
 \eqno{(7b)}
 $$
 $$
 \ddot{B}+{\dot{B}}^{2}=  {\omega\over{2}}{\phi^{\prime 2}\over{\phi^{2}}}
 +{\phi^{\prime\prime}\over{\phi}}.
 \eqno{(7c)}
 $$

 The wave equation (2) can be written as
 $$
 {\phi^{\prime\prime}\over{\phi}}+{\phi^{\prime}\over{\phi}}{C^{\prime}\over{C}}
 =-{2\sigma\over{(2\omega+3)\phi}}
 \eqno{(7d)}
 $$

 From (7b) and (7c) one can write
 $$
 \ddot{B}+{\dot{B}}^{2}=  -{\omega\over{2}}{\phi^{\prime 2}\over{\phi^{2}}}
 +{\phi^{\prime}\over{\phi}}{C^{\prime}\over{C}}=
 {\omega\over{2}}{\phi^{\prime 2}\over{\phi^{2}}}
 +{\phi^{\prime\prime}\over{\phi}}
 =b_{0}
 \eqno{(8)}
 $$
 where $b_{0}$ is a separation constant.

 From (8) one gets 
 $$
 {\phi^{\prime\prime}\over{\phi}}
 +{\phi^{\prime}\over{\phi}}{C^{\prime}\over{C}}
 =2 b_{0}
 \eqno{(9)}
 $$

 Hence the equation to be solved are
 $$
 {C^{\prime\prime}\over{C}}=-{\sigma\over{\phi}}-{\omega\over{2}}
 {\phi^{\prime 2}\over{\phi^{2}}}-2b_{0}
 \eqno{(10a)}
 $$
 $$
 {\phi^{\prime}\over{\phi}}{C^{\prime}\over{C}}-{\omega\over{2}}
 {\phi^{\prime 2}\over{\phi^{2}}}=b_{0}
 \eqno{(10b)}
 $$
 $$
 {\phi^{\prime\prime}\over{\phi}}+{\phi^{\prime}\over{\phi}}{C^{\prime}\over{C}}
 =2b_{0}
 \eqno{(10c)}
 $$
 $$
 {\sigma\over{(2\omega+3)\phi}}=-b_{0}
 \eqno{(10d)}
 $$
 $$
 \ddot{B}+{\dot{B}}^{2}=b_{0}
 \eqno{(10e)}
 $$
 Here we have four unknowns $C, \phi, \sigma, B$ and we have five equations.
 However one can show that (10a) is not independent equation but can be 
 obtained from rest of the equations.

 In what follows we shall try to solve the system of equations for 
 $b_{0}=0$ and for $b_{0}\neq 0$.

 {\bf{CASE I :}} {\bf{\underline{$b_0$=0}}}

 In this case, from (10e), one can solve $B$ which becomes
 $$
 e^{B} = t
 \eqno{(11)}
 $$
 where the constants of integration have been absorbed by rescaling the time
 coordinate without any loss of generality.\\
 A coordinate transformation of the form \cite{R12}
 $\xi=t Sinhz$ and $\tau=tCoshz$\\
 reduces this solution to the static form. That is the case $b_{0}=0$ 
 essentially represents a static case.\\
 From (10d), however, one can see that either $\sigma=0$ for finite value
 of $\omega$, or $\omega \rightarrow \infty$ for finite value of $\sigma$.
 But we know that for large value of $\omega$, the B-D theory is 
 indistinguishable from GR when $T=T^{\mu}_{\mu}\neq 0$ \cite{R13}. Hence
 we cannot have a cosmic string with $T^{\mu}_{\nu}$ given by (4a) and (4b)
 in B-D theory for $b_{0}=0$ which is a static case. This result is 
 completely on agreement with the previous work \cite{R8} that a static local
 gauge string of Vilenkin type does not exist in B-D theory of gravity.
\newpage

 {\bf{CASE II:}} {\underline{$b_0 \neq 0$}}

 In this case, one should note from eqn (10d) that if $\phi>0$, which 
 ensures the positivity of $G$, the gravitational constant, and $(2\omega+3)$
 is also positive, then to have positive energy density $\sigma$, $b_{0}$ should be negative
 and hence we take
 $$
 b_{0}= -b_{1}^{2}
 \eqno{(12)}
 $$
 where $b_{1}$ is a real constant.

 It deserves mention at this point that Gregory \cite{R14} found a nonsingular
 solution for a global string in GR for nonstatic case. For that, one has to 
 take a positive value of the constant $b_{0}$, and even for the global string,
 the energy density corresponding to the string ($T^{0}_{0}$, to be precise)
 takes a negative value for a certain value of the radial distance from the axis
 of the string.

 Now from (10e), we have
 $$
 {\ddot{(e^{B})}\over{e^{B}}}= -b_{1}^{2}
 $$
 which on integration yields
 $$
 e^{B}= MSin(b_{1}t)+NCos(b_{1}t)
 \eqno{(13)}
 $$
 where $M$ and $N$ are constants of integration.\\
 From (10b) and (10c) one gets,
 $$ 
 {\phi^{\prime\prime}\over{\phi}}+{\omega\over{2}}
 {\phi^{\prime 2}\over{\phi^{2}}} = -b_{1}^{2}.
 \eqno{(14)}
 $$
 Solving eqn (14), one gets,
 $$
 \phi=\phi_{0}[Cos(b_{1}ar)]^{1/a^2}
 \eqno{(15)}
 $$
 where $a^{2}=(\omega+2)/2$ and $\phi_{0}$ is a constant of integration.

 Again from (10b) and (10c) one gets,
 $$
 {\phi^{\prime\prime}\over{\phi}}+{\omega}
 {\phi^{\prime 2}\over{\phi^{2}}} = {\phi^{\prime}\over{\phi}}
 {C^{\prime}\over{C}}
 \eqno{(16)}
 $$
 Now if one assumes $\textstyle{{\phi^{\prime}\over{\phi}}}\neq 0$, 
 then one can integrate the above equation to yield
 $$
 C=K\phi^{\prime}\phi^{\omega}
 \eqno{(17)}
 $$
 where $K$ is a constant of integration.

 Using (15) and (16) one gets,
 $$
 C^{2}=C_{0}^{2}Sin^{2}(b_{1}ar)[Cos(b_{1}ar)]^{2\omega/(\omega+2)}
 \eqno{(18)}
 $$
 where $C_{0}^{2}= \textstyle{{2b_{1}^{2}K^{2}\phi_{0}^{2(\omega+1)}
 \over{(\omega+2)}}}.$

 The string energy density $\sigma$ can be found from (10d) which
 becomes
 $$
 \sigma = \phi_{0}b_{1}^{2}(2\omega+3)[Cos(b_{1}ar)]^{2/(\omega+2)}
 \eqno{(19)}
 $$

 Finally the line element becomes
 $$
 ds^{2}=dt^{2}-dr^{2}-[MSin(b_{1}t)+NCos(b_{1}t)]^{2}dz^{2}-
 C_{0}^{2}Sin^{2}(b_{1}ar)[Cos(b_{1}ar)]^{2\omega/(\omega+2)}d\theta^{2}
 \eqno{(20)}
 $$
 One should note that for $r \rightarrow 0$ i.e. near the axis of the string,
 the line element becomes,
 $$
 ds^{2}=dt^{2}-dr^{2}-[MSin(b_{1}t)+NCos(b_{1}t)]^{2}dz^{2}-
 C_{0}^{2}b_{1}^{2}a^{2}r^{2}d\theta^{2}
 \eqno{(21)}
 $$

 As a conclusion, one can say that a consistent set of interior solutions
 of the Einstein's equations for a Vilenkin type local gauge string can be
 obtained in B-D theory if it is nonstatic, although a static string of this
 type is inconsistent in this theory.

 The nonstatic metric (20) shows that the proper volume becomes zero 
 periodically in time and hence there is disk like singularity in spacetime
 periodically with time. But this fact is not reflected in curvature scalar
 given by
 $$
 R=2\omega b_{1}^{2}[{Tan^{2}(b_{1}ar)\over{(\omega+2)}}-{2\over{(2\omega+3)}}]
 \eqno{(22)}
 $$
 which is evidently time independent. This result is similar to the case for
 nonstatic global string both in GR and in dilaton gravity earlier obtained
 by Gregory et.al \cite{R14}\cite{R10}. There also it can be shown that for
 $b_{0}<0$, the proper volume periodically vanishes with time, whereas the
 curvature scalar does not. It can also be seen  from (22) that the space time
 becomes singular periodically at finite distances from the axis of the string.
 \newpage
 {\bf{ACKNOWLEDGEMENT}}

 The author is grateful to University Grants Commission, 
 India for financial support.


\begin{thebibliography}{30}

 \bibitem{R1}C.Gundlach and M.E.Ortiz, Phys.Rev.D, {\bf{42}},2521 (1990).
 \bibitem{R2}A.Barros and C.Romero, J.Math.Phys, {\bf{36}}, 5800 (1995).
 \bibitem{R3}M.E.X Guimar\~{a}es, Class.Quantum.Grav, {\bf{14}}, 435 (1997).
 \bibitem{R4}T.Damour and A.Vilenkin, Phys.Rev.Lett, {\bf{78}},2288 (1997).
 \bibitem{R5}R.Gregory and C.Santos, Phys.Rev.D, {\bf{56}}, 1194 (1997).
 \bibitem{R6}B.Boisseau and B.Linet, gr-qc/9802007.
 \bibitem{R7}A.Vilenkin, Phys.Rep, {\bf{121}}, 263 (1985).
 \bibitem{R8}A.A.Sen, N.Banerjee and A.Banerjee,Phys.Rev.D, {\bf{53}}, 5508 (1997).
 \bibitem{R9}A.A.Sen and N.Banerjee, Phys.Rev.D (in press), gr-qc/9711036.
 \bibitem{R10}O.Dando and R.Gregory, gr-qc/9802015.
 \bibitem{R11}J.A.Stein-Schabes, Phys.Rev.D, {\bf{33}},3545, (1986).
 \bibitem{R12}L.D.Landau and E.M.Lifshitz, {\bf{Classical thoery of fields}},
 (Pergamon, Oxford, 1962).
 \bibitem{R13}N.Banerjee and S.Sen, Phys.Rev.D {\bf{56}}, 1334 (1997).
 \bibitem{R14}R.Gregory, Phys.Rev.D, {\bf{54}}, 4955 (1996).
 \end{thebibliography}
\end{document}